\documentclass[draft]{agujournal2019}

\usepackage[utf8]{inputenc}
\usepackage[english]{babel}
\usepackage{amsmath}
\usepackage{tabularx}
\usepackage{array}
\usepackage{natbib}
\usepackage{comment}
\usepackage{placeins}
\usepackage{subcaption}
\usepackage{ragged2e}
\usepackage{soul}

\justifying

\draftfalse

\newcolumntype{Y}{>{\centering\arraybackslash}X}

\RequirePackage[a4paper,portrait]{geometry}
\journalname{JGR: Space Physics}

\begin{document}
    \title{Ionospheric conductances \\
    at the giant planets of the Solar System:
    \\
    a comparative study of ionization sources \\
    and the impact of meteoric ions}
    \authors{
    No\'e Cl\'ement\affil{1, 2},
    Yuki Nakamura\affil{3},
    Michel Blanc\affil{2, 4},
    Yuxian Wang\affil{5},
    Sariah Al Saati\affil{2, 6}
    }
    \affiliation{1}{Laboratoire d'Astrophysique de Bordeaux, Univ. Bordeaux, CNRS, B18N, allée Geoffroy Saint-Hilaire, 33615 Pessac, France}
    \affiliation{2}{IRAP, CNRS, Université Toulouse III-Paul Sabatier, CNES, Toulouse, France}
    \affiliation{3}{Graduate School of Science, The University of Tokyo, Tokyo, Japan}
    \affiliation{4}{Laboratoire d'Astrophysique de Marseille, Aix-Marseille-Université (AMU), CNRS, Marseille, France}
    \affiliation{5}{State Key Laboratory of Space Weather, National Space Science Center, Chinese Academy of Sciences, Beijing, China}
    \affiliation{6}{CPHT, CNRS, Institut Polytechnique de Paris, Palaiseau, France}
    
    \vspace{2cm}
    \begin{center}
    Submitted July, 12$^\text{th}$ 2024 - Accepted December, 5$^\text{th}$ 2024
    \end{center}

    \correspondingauthor{Michel Blanc}{michel.blanc@irap.omp.eu}
    \correspondingauthor{No\'e Cl\'ement}{noe.clement@u-bordeaux.fr}
    
    \begin{keypoints}
        \item We estimate the contributions of the main ionization sources, including the meteoric ions, to the Hall and Pedersen conductances.
        \item At Saturn, Uranus, and Neptune, the contribution of meteoric ions to conductances could be non-negligible.
        \item At these three planets, the layer where meteoric ions are produced is deeper than the conductive layer, limiting their role in M-I coupling.
    \end{keypoints}

    \begin{abstract}
    The dynamics of giant planet magnetospheres is controlled by a complex interplay between their fast rotation, their interaction with the solar wind, and their diverse internal plasma and momentum sources. In the ionosphere, the Hall and Pedersen conductances are two key parameters that regulate the intensity of currents coupling the magnetosphere and the ionosphere, and the rate of angular momentum transfer and power carried by these currents.
    We perform a comparative study of Hall and Pedersen conductivities and conductances in the four giant planets of our Solar System - Jupiter, Saturn, Uranus and Neptune.
    We use a generic ionospheric model (restraining the studied ions to H$_3^+$, CH$_5^+$, and meteoric ions) to study the dependence of conductances on the structure and composition of these planets' upper atmospheres and on the main ionization sources (photoionization, ionization by precipitating electrons, and meteoroid ablation).
    After checking that our model reproduces the conclusions of \citet{Nakamura2022} at Jupiter, i.e. the contribution of meteoric ions to the height-integrated conductances is non-negligible,
    we show that this contribution could also be non-negligible at Saturn, Uranus and Neptune, compared with ionization processes caused by precipitating electrons of energies lower than a few keV (typical energies on these planets). However, because of their weaker magnetic field, the conductive layer of these planets is higher than the layer where meteoric ions are mainly produced, limiting their role in magnetosphere-ionosphere coupling.
    \end{abstract}

%%%%%%%%%%%%%%%%%%%%%
% 1 - introduction
%%%%%%%%%%%%%%%%%%%%%

\section{Introduction}
\subsection{Ionospheric conductances:  mediators between the magnetosphere and the rotating planet}

Hall and Pedersen conductivity profiles and their associated height-integrated conductances are the key parameters that characterize how much the ionosphere connects a rotating planet to its magnetosphere by transporting electric charges.
The Pedersen conductivity drives electric charge transport mainly perpendicularly to the aurora and connects the downward field-aligned currents to the upward field-aligned currents, thus closing the electrical circuit. The Hall conductivity rules electric charge transport mainly along the aurora.
The Hall and Pedersen conductivities result from the ionization of the atmosphere. Their intensities are linear functions of ion and electron densities and reach their maximum in the regions where the cyclotron frequencies of ionospheric charged particles are of the same order of magnitude as their collision frequencies with atmospheric neutral species, thus maximizing the mobility of electric charges perpendicular to the magnetic field \citep{Banks1973}.
\par
Since conductances are linear functions of ion and electron densities, their determination starts from an assessment of the different ionization sources of neutral species in planetary upper atmospheres, which are:
\begin{itemize}
    \item photoionization by solar and stellar fluxes;
    \item inelastic collisions of precipitating electrons and ions with neutral particles, typically for incoming particle energies of 30 eV and more, which strip electrons off neutral atoms and molecules (this source is generally restricted to high-latitude and auroral regions);
    \item ionization by the precipitation of solid bodies, such as meteoroids and planetary/interplanetary dust \citep{Nakamura2022};
    \item ionization by galactic cosmic rays \citep{MolinaCuberos2023}, neglected in this study.
\end{itemize}

\par
The narrow outer layers of the atmosphere of giant planets where conductivities are at their maximum (and contribute to most of these planets' conductances) play a central role in the coupling of these atmospheres with the two main external momentum sources that drive their magnetosphere into motion: (1) the solar wind; (2) giant planets moons. 
%%%\color{red}
In the case of weakly magnetized bodies, such as Venus and Mars for instance, whose atmosphere directly interacts with the solar wind flow, \citet{Vernisse2017} showed that the intensity of the ionospheric conductance of these atmospheres can be used to characterize the strength of the obstacle the atmosphere opposes to the solar wind and to classify these types of interactions into different dynamical regimes. In this article, we will focus on the case of giant planet magnetospheres, within which atmospheres do not directly interact with the solar wind flow. 
%%%\color{black}
When magnetospheric field lines are set into motion and when plasma flows across these field lines can be regarded as in a steady state, the divergence of electric currents generated by these flows across field lines results in local charge accumulation. Since electric currents have to be divergence-free in a steady state, this accumulation of charges drives current systems that preferentially close along the paths of highest conductivity in the system. First, currents flow between the magnetosphere and the ionosphere along magnetic field lines, enabled by the very high mobility of low-energy electrons along them, forming what are called ``field-aligned currents" (FAC's), also called Birkeland currents. Then, they close across field lines in the only layer where significant transverse conductivities exist, i.e. in the planet’s upper atmosphere near the altitudes where $\nu_{\text{in,en}}/\omega_{\text{ci,ce}}=1$ ($\nu_{\text{in,en}}$ being the collision frequency of ions with neutrals (i-n) or electrons with neutrals (e-n), and $\omega_{\text{ci,ce}}$ the cyclotron frequency of ions (ci) or electrons (ce)).
%%%\color{red}
This altitude domain, where conductivities reach their maximum intensities, is usually called the ionosphere dynamo layer \citep{Riousset2013, Riousset2014}.
% “The altitude domain where conductivities reach their maximum intensities, around the altitudes where the ratios of charged particle collision frequencies with the neutral gas to their gyrofrequencies are of the order of unity (see equations (2) and (3)) is usually called the ionosphere dynamo layer"
%%%\color{black}

\par
By this token, any charge locally accumulated by a partly divergent magnetospheric current system is discharged by a current system connecting currents in the conducting layer of the upper atmosphere to pairs of upward and downward field-aligned currents. The full current loop resulting from this process transfers momentum between its magnetospheric and ionospheric ends, via the corresponding Laplace force per unit volume (JxB) acting on the plasma. At the magnetospheric end, this JxB force accelerates magnetospheric plasma motions. At the ionospheric end, the corresponding JxB force is balanced by frictions between the ionospheric plasma and the background neutral air, and produces a net exchange of momentum between the neutral and charged particle components of the upper atmosphere.

\par
\begin{figure}[ht!]
    \centering
    \includegraphics[width=\textwidth]{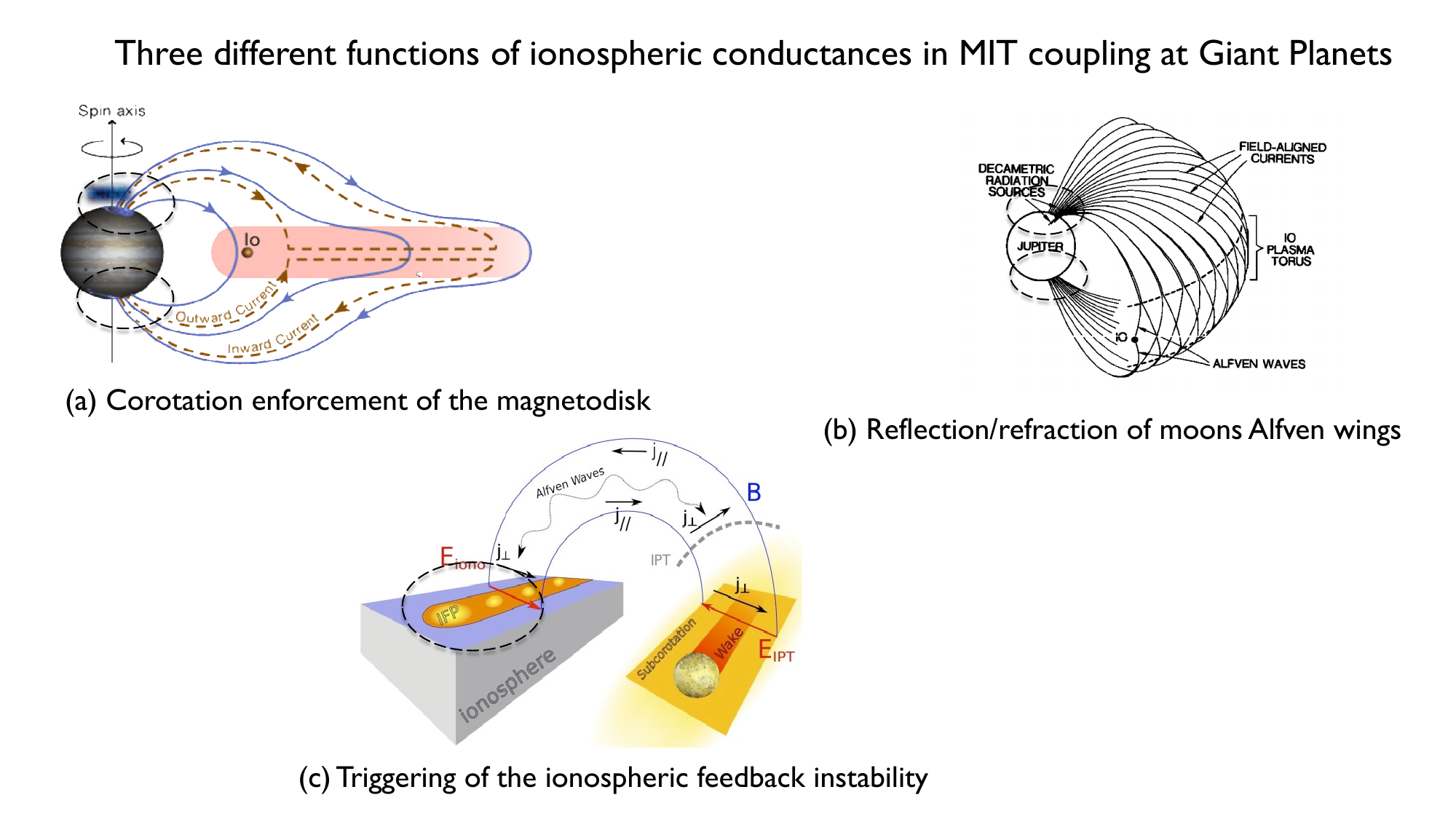}
    \caption{An illustration of the important roles of Pedersen and Hall conductances in magnetosphere-ionosphere coupling at giant planets, with emphasis on the case of Jupiter: (a) The Pedersen conductance plays a key role in the closing of the current system that connects Jupiter’s upper atmosphere to its magnetodisk and in the efficiency of enforcement of corotation on the magnetodisk \citep{Cowley2001}; (b) Ionospheric conductances determine the reflection and transmission properties of the thin conducting ionospheric layer along the train of multiple Alfven wing reflections that is generated by Io’s interaction with the corotating magnetospheric plasma (from \citet{Gurnett1981}) (c) The modulation of ionospheric conductances by precipitating charged particles, triggering the so-called ionospheric feedback instability, may be the cause of the generation of small-scale fluctuations in H$_3^+$ auroral brightness along moon auroral tails \citep{Moirano2021}.}
    \label{fig:3_examples_role_conductances}
\end{figure}

The most important example of this transfer of momentum by steady state current loops in giant planets' magnetospheres is the coupling of the plasmasheets/magnetodisks of Jupiter and Saturn to the rotating ionospheres/thermospheres of these planets, and its role in the corotation of these disks with their host planet, illustrated in Figure \ref{fig:3_examples_role_conductances}a. As first shown by \citet{Hill1979}, later developed by \citet{Cowley2001} and follow-on publications, and recently revisited by \citet{Devinat2024}, in the presence of an internal source injecting fresh plasma deep inside the magnetosphere at a rate of $\dot{M}$ (kg/s), the plasma disk generated by the outward diffusion or convection of this plasma is maintained into cororation up to a characteristic distance, the so-called cororation radius $R_c$, which scales as:

\begin{equation}
R_\text{c} = R_\text{p} \left(\frac{\Sigma_\text{P} \pi  R_\text{p}^2 B_\text{p}^2}{\dot{M}}\right)^{\frac{1}{4}}
\end{equation}

%%%\color{red}
\noindent where $R_\text{p}$ (m) is the planetary radius, $\Sigma_\text{P}$ (mho) the Pedersen conductance, $B_\text{p}$ (T) the strength of the dipolar magnetic field at the planet's surface, and $\dot{M}$ (kg/s) is the total mass loading rate of fresh plasma injected into the magnetodisk per unit of time by moon sources, as per equation (2) of \citet{Hill1979}.
%%%\color{black}

\par
As the equation shows, the corotation radius varies as the power 1/4 of the Pedersen conductance for a given value of $\dot{M}$: the larger the Pedersen conductance, the larger is the radial distance up to which the magnetodisk can be regarded as rigidly rotating with the planet.
Conversely, when the characteristic time scale of a magnetospheric plasma flow is shorter than the Alfven travel time along magnetic field lines between the location of the flow and the ionosphere, a steady state cannot be assumed anymore. Momentum exchange between magnetosphere and ionosphere is no longer carried by steady-state current systems, but instead by torsional MHD waves, i.e. Alfven waves, propagating along field lines. When plasma is set into motion at one location of the field line by a local momentum source (planetary rotation, solar wind interaction, interaction of the plasma flow with a moon obstacle, mass loading of the plasma flow by injection of freshly ionized neutrals…), momentum is transferred by Alfven waves propagating along field lines in the two opposite directions. When they reach the ionosphere, incident Alfven waves experience a complex interaction with this conducting layer, described in detail in the ideal MHD approximation by \citet{Nishida1978}: they are partly reflected back by the ionospheric conductor, partly refracted through it, generating electromagnetic waves in the underlying neutral atmosphere, and partly converted to a different MHD mode, preferentially the fast MHD mode in that case. The energy and momentum fluxes of the incident wave are partly redistributed between the three outgoing waves, and partly dissipated into the ionospheric conductor. When the wavelength of the incident Alfven wave is larger than the thickness of the ionospheric conductor, it can be treated as infinitely thin, and the properties of reflected, transmitted and converted waves can be fully characterized by the integral of Pedersen and Hall conductivities across the conducting layer, i.e. the Pedersen and Hall conductances, and by the “Alfven conductance” 1/($\mu_0 v_A$) of the incident wave, $v_A$ being the Alfven velocity. More recent calculations of this interaction, including kinetic effects and the large variations of magnetic field intensity along field lines \citep{Lysak2003,Saur2018} added new complexities to this simplified picture, such as the production of energetic electron beams accelerated by the field-aligned component of the wave electric field and their precipitation into the conjugate ionospheres which amplify conductances. The important role played by ionospheric conductances remains unchanged. 

\par
The most studied example of this transfer of momentum by Alfven waves is provided by the interactions of medium and large size regular moons of Jupiter and Saturn with their host magnetosphere and planet, illustrated in Figure \ref{fig:3_examples_role_conductances}b with the case of Io. As shown in the figure, the interaction of the magnetospheric plasma flow with Io produces a local deviation of this flow around the moon obstacle and drives the emission of a special train of Alfven waves called Alfven wings, as described theoretically by \citet{Goertz1980} and \citet{Neubauer1980}. The first observation of this interaction by Voyager I was described by \citet{Acuna1981}. Looking at a larger scale encompassing both moon and planet, one can see that the Alfven wing generated at the location of Io produces multiple bounces between the two conjugate ionospheres downstream of Io along the mainly corotating magnetospheric plasma flow. At each of these ionospheric reflections, a fraction of the momentum and energy carried by the Alfven wing is deposited into the upper atmosphere of Jupiter via the interaction of the Alfven wing with the conducting ionosphere. The amount of this deposition is directly determined by the values of the Pedersen and Hall conductances at and around the reflection area of the Alfven wing. 

\par
In both cases (steady-state current system or Alfven waves), the Hall conductance, i.e. the anisotropic properties of the magnetized ionospheric plasma, plays an important role that should not be neglected in addition to the Pedersen conductance. While Pedersen conductance plays the main role when conductances are spatially uniform, Hall conductances become important in the presence of horizontal variations in ionospheric properties, which is the general case. Indeed, in the case of horizontally uniform conductances, Hall currents are approximately divergence-free, since they flow and close along equipotentials of the electrostatic field without producing charge accumulation. However, as soon as conductances vary in the ionosphere along these equipotentials, Hall currents also tend to diverge. They contribute to the formation of field-aligned currents, and therefore to magnetosphere/ionosphere coupling (MI coupling for short) similarly to Pedersen currents. Hence the practical importance of knowing both conductances to correctly represent MI coupling in models and in the interpretation of observations.

\par
This contribution of Hall conductances to MI coupling becomes particularly important in the presence of electron precipitation, as it happens in high magnetic latitude regions of giant planet magnetospheres where they are responsible for a significant and often dominant fraction of upper atmosphere ionization. At these latitudes,  charged particle precipitation generates bands of enhanced conductances, as well as large conductance gradients at their equatorward and poleward edges. The interaction of these strong electron precipitation regions with the closure of field-aligned currents in the ionosphere has long been documented in depth at Earth where they play a dominant role in the determination of the horizontal distribution of conductances \citep{Fontaine1983} and of the general pattern of plasma convection in response to solar wind forcing \citep{Senior1984, Fontaine1985}. It also plays a major role in the determination of ionospheric plasma flow patterns at giant planets. \citet{Nichols2004} showed that the amplification of conductances by electron precipitation at the feet of the ionosphere/magnetosphere current loop (Figure \ref{fig:3_examples_role_conductances}a) concentrates the entry of field-aligned currents in this region of high conductances and reduces the latitudinal width of the main auroral oval emission in comparison with predictions based on spatially uniform conductances. 

\par
Finally, the modulation of ionospheric conductances by charged particle precipitation transforms the ionosphere from a purely resistive medium into an active medium that can drive instabilities and generate small-scale structures in MI coupling current systems. Indeed, Pedersen and Hall conductances are functions of the energy flux and characteristic energy of precipitating electrons. Since precipitation fluxes contribute themselves to the net current flow along field lines (though in a complex and still poorly understood way), the resistive properties of the ionospheric conductor become functions of the current flowing through it, just like in active electronic circuits. This non-linear relationship between field-aligned currents and ionospheric conductances has profound effects on the dynamical properties of MI coupling. \citet{Miura1980}, \citet{Lysak1991}, \citet{Lysak2002}, \citet{Watanabe2010}, \citet{Hiraki2015}, \citet{Streltsov2018} and other authors showed how these properties can generate small-scale auroral structures (auroral arcs, curls and other spectacular features of auroral displays) via a feedback response of the ionosphere to magnetospheric currents and MHD waves. 

\par
This so-called “ionospheric feedback instability” has also been found to play an important role in the formation of the moons auroral tails produced by moons-magnetosphere coupling at giant planets, which is studied in considerable detail with NASA’s Juno mission to Jupiter. \citet{Moirano2021} observed small-scale intensity variations in IR emissions along moon auroral tails at scales of a few 100 km in Juno’s JIRAM infra-red spectral images. These fluctuations develop at smaller scales than the distance separating successive reflections of the moon Alfven wings, and are stationary with respect to the Jovian atmosphere rather than to the reference frame of moons. They proposed that this mechanism results from a type of ionosphere feedback instability acting at Jupiter (\ref{fig:3_examples_role_conductances}c). Again, determining the variations of Pedersen and Hall conductances along and across moon auroral tails will be critical to distinguish between the different possible generation mechanisms of these tails and of their intriguing sub-structures. 

\subsection{Ionospheres of giant planets in the Solar System}

The atmospheres of the four giant planets of our Solar System all display a chemical composition dominated by molecular hydrogen and helium. Methane is the most abundant hydrocarbon species at the four giants.
They display the same type of troposphere-stratosphere-thermosphere thermal structure, and their ionospheres extend from the top of their thermospheres down to their upper stratospheres.
These similarities allow us to compare their ionospheric conductivities, making the following assumption: we will only consider H$_2$ and its associated ions H$_{2}^{+}$ and H$_{3}^{+}$, CH$_4$ and its associated ion CH$_{5}^{+}$, and the chemical reactions involving these species. We will then add meteoric ions.
At Jupiter and Saturn, H$_3^+$ has been detected and documented for a long time \citep{Drossart1989, Moore2019}. It has also been detected at Uranus \citep{Trafton1993, Melin2020, Thomas2023}. At Neptune, the H$_3^+$ ion and aurora remain undetected \citep{Moore2020}. However, the recent detection of an infrared aurora at Uranus and the associated increase in H$_3^+$ emissions suggest that Neptune may exhibit similar phenomena. In this study, we will assume that the H$_3^+$ ion exists in Neptune's upper atmosphere.
Our parameter space can be summarized as follows:
\begin{itemize}
    \item Magnetic field intensities
    \item Temperature profiles
    \item Density profiles of neutral species - H$_2$ and CH$_4$ - which are controlled by gravity field intensities
\end{itemize}

\par
The diversity of the four giants' parameter field could even be enriched by exoplanets.
Solar System giant planets are the templates of a large population of giant exoplanets displaying similar masses that have been identified in our galactic neighborhood: ``hot" or ``warm" Jupiters, and Neptune-like planets, which likely share similar atmospheric structures with them. Lessons learned at giant planets could partly be applied to this population of exoplanets.

\par
In this study, we compare ionospheric conductivities and conductances in the four giant planets and the contribution of meteoric ions to these parameters, with a simplified model that can be applied to all these planets.
In Section \ref{Method}, we present this model and its assumptions. In Section \ref{Conductivities} we look at the conductivity profiles and in Section \ref{Conductances} at the conductances that we calculate with this model, highlighting the impact of meteoric ions on both. In Section \ref{Altitudes} we analyze the implications of our results on magnetosphere-ionosphere coupling.
Finally in Section \ref{Discussion}, we discuss the limitations of our modeling and open questions.

%%%%%%%%%%%%%%%%%%%%%
% 2 - method
%%%%%%%%%%%%%%%%%%%%%

\section{A generic ionospheric model for giant planets}
\label{Method}

In this section, we present our versatile and generic model and the approximations that we have made to build it in order to compare the four giant planets.
\par
\begin{figure}[ht!]
    \centering
    \includegraphics[width=\textwidth]{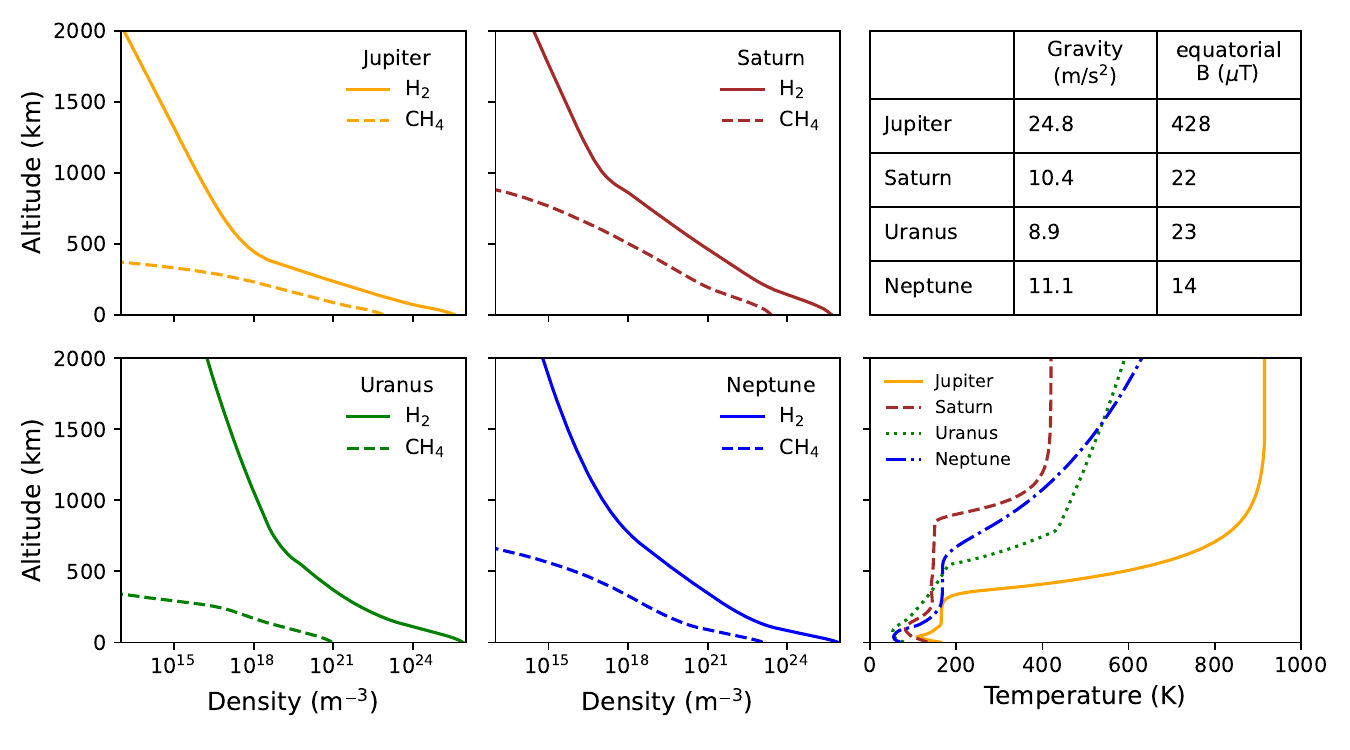}
    \caption{An illustration of the neutral atmosphere component of our parameter space: H$_2$ and CH$_4$ densities, temperature profiles from \citet{Moses2017} as functions of altitude, equatorial magnetic field and gravity. While gravity does not appear explicitly, together with the neutral gas temperature it controls the vertical extent of each atmosphere.}
    \label{fig:parameter_field}
\end{figure}

Our model aims to estimate the altitude distributions of Pedersen and Hall conductivities, $\sigma_\text{P}$ and $\sigma_\text{H}$, in the auroral regions. The ionospheric conductances will then be retrieved by vertical integration of the conductivities.
To estimate the electron and ion densities needed to calculate Hall and Pedersen conductivities, we use the same ionospheric model as \citet{Wang2021} for Jupiter, which we adapt to the four giant planets and to which we add the ion production rates by photoionization and by meteoric sources \citep{Nakamura2022}.
In this generic model, the only neutral species that we consider are H$_2$ and CH$_4$ (see their vertical density profiles in Figure \ref{fig:parameter_field}).
H$_2$ represents about 80\% of the atmospheric mass density at ionospheric altitudes, so only H$_2$ will be considered to contribute to ion-neutral collisions. By neglecting He and H, we estimate that this approximation would not alter the collision frequencies by more than 20\%.
CH$_4$ is also included in the model because it is the main species (and the main hydrocarbon), reacting with the dominant ion H$_3^+$.
\citet{Hiraki2008} showed that there is little dependence of the estimated ionospheric conductances on atomic species such as H and He.
%%%\color{red}
Only solar sources are taken into account in the photoionization process. Extra-solar sources have little impact and are not taken into account, as can be validated by the results of previous ionosphere equilibrium models that included both solar and stellar ionization sources. Quantitatively, extrapolating to the heliocentric distances of the four giant planets the model calculations of main ion production rates performed for Earth by \citet{Lei2004} and previously by \citet{Titheridge2000} who both considered a daytime solar source and a nighttime stellar source, shows that the stellar ionization rate is about 200 to 300 times smaller than the solar one at Jupiter, and still 10 times smaller than the solar one at Neptune. Since this study focuses on the contribution of meteorites anyway, neglecting the stellar flux for all giant planets is a reasonable assumption.
%We estimate that the ionization rate they induce does not exceed 10\% of the ionization rate induced by the solar flux, at the position of Neptune, i.e. where the solar flux is weakest.
%%%\color{black}

\par
The reaction chain starts with
H$_2$ being ionized by precipitating electrons and photoionization to give H$_2^+$, which quickly reacts with H$_2$ to produce H$_3^+$  and H (\citet{Majeed1991}, see Table \ref{tab:reactions}).
H$_3^+$ ions then react with CH$_4$ to give CH$_5^+$.
We assume that CH$_5^+$ ions are the final products of ion-neutral chemical reactions.
%%%\color{red}
Photoionization of CH$_4$ is not taken into account.
To check the impact of this assumption, we calculated the photoionization rates of CH$_4$, which are always at least one order of magnitude lower than the production rates of CH$_5^+$ by the (CH$_4$ + H$_3^+$ $\rightarrow$ CH$_5^+$ + H$_2$) reaction.
%%%\color{black}
In the first step, the only ions we consider contributing to conductivities are H$_3^+$ and CH$_5^+$.
This simplification allows us to compare the four giant planets.
In the second step, we add the main ions produced by meteoroid precipitation and ablation - Fe$^+$, Mg$^+$, Si$^+$, and Na$^+$ - to estimate their contribution to conductivities.

The conductivities $\sigma_\text{P}$ and $\sigma_\text{H}$ are calculated as follows:

\begin{equation}
    \sigma_\text{P} = \dfrac{n_\text{e} e}{B} \dfrac{\left|\omega_\text{ce}\right|\nu_\text{en}}{\nu_\text{en}^2 + \omega_\text{ce}^2} + \sum_{\text{ions}} \dfrac{n_\text{i} e}{B} \dfrac{\left|\omega_\text{ci}\right|\nu_\text{in}}{\nu_\text{in}^2 + \omega_\text{ci}^2}
    \label{sigma_P}
\end{equation}

\begin{equation}
    \sigma_\text{H} = \dfrac{n_\text{e} e}{B} \dfrac{\omega_\text{ce}^2}{\nu_\text{en}^2 + \omega_\text{ce}^2} - \sum_{\text{ions}} \dfrac{n_\text{i} e}{B} \dfrac{\omega_\text{ci}^2}{\nu_\text{in}^2 + \omega_\text{ci}^2}
    \label{sigma_H}
\end{equation}

\noindent where $n_\text{e}$ and $n_\text{i}$ are electron and ion number densities respectively, $e$ is the elementary charge, $\omega_\text{ce}$ and $\omega_\text{ci}$ are the cyclotron frequencies of electrons and ions, $B$ is the local magnetic field. Because this model is designed to be applied in the polar magnetic regions (the auroral regions), the value used for the magnetic field is the polar value. To calculate it, we multiply by 2 the equatorial magnetic field, by assuming that the magnetic fields of the giant planets are dipoles at first approximation. $\nu_\text{en}$ and $\nu_\text{in}$ are the collision frequencies between electrons/neutrals and ions/neutrals taken from \citet{Banks1973}:

\begin{equation}
\nu_\text{in} (\text{s}^{-1}) = n_{\text{H}_2} (\text{cm}^{-3}) \times 2.6 \times 10^{-9} \times \sqrt{\frac{\alpha_0}{\mu_\text{a}}} %/s
\label{i_n_collision_frequency}
\end{equation}

\begin{equation}
\nu_\text{en} (\text{s}^{-1}) = 2 \times n_{\text{H}_2} (\text{cm}^{-3}) \times 2.5 \times 10^{-9} \times (1 - 1.35 \times 10^{-4} \times T) \times T^{1/2}
\end{equation}

\noindent where $n_{\text{H}_2}$ and $T$ are density and temperature profiles (Figure \ref{fig:parameter_field}), $\alpha_0=0.82$ is the polarizability of H$_2$ and $\mu_\text{a}$ the reduced mass in atomic mass unit (AMU) for a H$_2$-ion pair where the ion is one of those considered in our model.
The masses of the ions we consider vary from 3 AMU for H$_3^+$ to 56 AMU for Fe$^+$, the reduced mass for a H$_2$-ion pair then varies from 1.2 AMU to 1.93 AMU. For $\nu_\text{en}$, we have multiplied by 2 the expression given by \citet{Banks1973} for H, estimating that the electron momentum transfer cross section of H$_2$ should be about twice the one of H.

\noindent \textbf{$\text{H}_2^+$, $\text{H}_3^+$, $\text{CH}_5^+$ production}.
\\
The calculation of $\text{H}_2^+$ production by electron precipitation relies on the parameterization of \citet{Hiraki2008} which aims to calculate ionization rates of $\text{H}_2^+$ in any H$_2$-dominated atmosphere.
%%%\color{red}
The differential ionization rate (m$^{-1}$) $q_\text{ie}(\epsilon_0,z)$ of H$_2$ at altitude $z$ generated by the precipitation of an auroral electron with an initial energy $\epsilon_0$ (eV) is given by:

\begin{equation}
    q_\text{ie}(\epsilon_0,z)= \frac{\epsilon_0}{\epsilon_\text{i}} \frac{\lambda(z)}{R_0(\epsilon_0)} \rho(z)
\end{equation}

\noindent where $\epsilon_\text{i}$ is the ion/electron pair yield energy (30 eV),
$\lambda(z)$ a non-dimensional parameter representing the normalized energy dissipation distribution function \citep{Rees1963},
$\rho(z)$ the mass density (kg/m$^3$) of neutral gas (H$_2$) and $R_0(\epsilon_0)$ the column mass density (kg/m$^2$) which is defined as the integration of $\rho$ above the stopping height of the electron, which is depending on $\epsilon_0$.
%%%\color{black}

For a given energy flux distribution $F(\epsilon)$ (W m$^{-2}$ eV$^{-1}$) of precipitating electrons of energy $\epsilon$ (eV), the production rate $Q_\text{ie}$ (m$^{-3}$ s$^{-1}$) at altitude $z$ is: 
\begin{equation}
    Q_\text{ie}(z)= \int q_\text{ie}(\epsilon,z) \frac{F(\epsilon)}{\epsilon} d\epsilon
    \label{flux_integral}
\end{equation}

For a mono-energetic beam F (W m$^{-2}$) of precipitating electrons of characteristic energy $\epsilon$ (eV), it becomes:
\begin{equation}
    Q_\text{ie}(z) = q_\text{ie}(\epsilon,z) \frac{F}{\epsilon}
    \label{flux_constant}
\end{equation}

To the production of $\text{H}_2^+$ by electron precipitation $Q_\text{ie}(z)$, we add the production of $\text{H}_2^+$ by photoionization $Q_\text{ip}(z)$, under averaged solar conditions, calculated with the model of \citet{Richards1994}, to obtain the total production of $\text{H}_2^+$ denoted $Q_\text{i}(z)$:

\begin{equation}
    Q_\text{i}(z) = Q_\text{ie}(z) + Q_\text{ip}(z)
\end{equation}

H$_2^+$ ions quickly transfer their charge to produce H$_3^+$ ions. $Q_\text{i}(z)$
%%%\color{red}
is in reasonable agreement with
%%%\color{black}
the production rate of $\text{H}_3^+$ \citep{Gerard2020}. 
Above the CH$_4$ homopause, $\text{H}_3^+$ recombine with electrons. Below it, H$_3^+$ can then react with CH$_4$ to produce CH$_5^+$ (reaction 5 in Table \ref{tab:reactions}), which can recombine with electrons.

\begin{table}[ht!]
\centering
\renewcommand{\arraystretch}{1}
\caption{Main chemical reactions considered in processes of photoionization and ionization by precipitating electrons.
%%%\color{red}
``Products" refer to neutral atomic and molecular species that do not contribute to the conductivities.
%%%\color{black}
}
\begin{tabularx}{\textwidth}{|l|X|X|}
    \hline
    reaction & rate coefficient (cm$^3$ s$^{-1}$) & reference \\
    \hline
    $\text{H}_{2} + \text{e}^{-} \rightarrow \text{H}_{2}^{+} + 2 \text{e}^{-}$ &
    ionization by precipitating electrons:
    calculated with the parameterization from \citet{Hiraki2008} & \citet{Hiraki2008} \\
    \hline
    $\text{H}_{2} + \nu \rightarrow \text{H}_{2}^{+} + \text{e}^{-}$ &
    photoionization:
    calculated with the model from \citet{Richards1994} & \citet{Richards1994} \\
    \hline
    $\text{H}_{2}^{+} + \text{H}_{2} \rightarrow \text{H}_{3}^{+} + \text{H}$ & dominant reaction implying H$_2^+$ as a reactant & \citet{Majeed1991,Gerard2020} \\
    \hline
    $\text{H}_{3}^{+} + \text{e}^{-} \rightarrow$ Products & $1.2 \times 10^{-7} (\frac{300K}{T_\text{e}})^{0.65}$ & \citet{Sundstrom1994}, \citet{Gerard2020}\\
    \hline
    $\text{H}_{3}^{+} +  \text{CH}_{4} \rightarrow  \text{CH}_{5}^{+}$ & $2.4 \times 10^{-9}$ & \citet{Perry1999} \\
    \hline
    $ \text{CH}_{5}^{+} + \text{e}^{-} \rightarrow$ Products & $2.7 \times 10^{-7} (\frac{300K}{T_\text{e}})^{0.52}$ & \citet{Perry1999}, \citet{Gerard2020}\\
    \hline    
\end{tabularx}
    \label{tab:reactions}
\end{table}

\noindent \textbf{Meteoric ion production}.
Production rates of meteoric ions Fe$^+$, Mg$^+$, Si$^+$, Na$^+$ are calculated with the meteoroid ablation model from \citet{Nakamura2022} at Jupiter adapted to the four giant planets. The runs of this model use the same atmospheric profiles from \citet{Moses2017} as the ionospheric model (Figure \ref{fig:parameter_field}). The flux, mass and speed distributions of the incident meteoroids are also taken from \citet{Moses2017}.
The element composition of the meteoroids is assumed to be CI-chondrite \citep{Kim2001}, in which the elements are Fe, Mg, Si, and Na with mass fractions of 19.0\%, 9.9\%, 10.6\%, and 0.5\%, respectively. Meteoric atoms ablated from the surface of the meteoroids collide with atmospheric molecules and are ionized to form meteoric ions. The ionization probability during this process depends on the entry velocity of meteoroids \citep{Lebedinets1973}, and it has been estimated to be 0.4 at Jupiter \citep{Kim2001}. We also use this value for the three other giant planets.

\begin{table}[ht!]
    \centering
    \renewcommand{\arraystretch}{1}
    \caption{Re-combinations of ions created by meteoric sources. Values are taken from \citet{Kim2001}.}
    \begin{tabular}[ht!]{|l|l|l|}
        \hline
        ion & reaction & rate coefficient (cm$^3$ s$^{-1}$) \\
        \hline
        Fe$^+$ & $\text{Fe}^+ + \text{e}^{-} \rightarrow$ Fe & $3.7 \times 10^{-7} (\frac{300K}{T_\text{e}})^{0.6}$ \\
        \hline
        Mg$^+$ & $\text{Mg}^+ + \text{e}^{-} \rightarrow$ Mg & $2.8 \times 10^{-7} (\frac{300K}{T_\text{e}})^{0.8}$ \\
        \hline
        Si$^+$ & $\text{Si}^+ + \text{e}^{-} \rightarrow$ Si & $4.9 \times 10^{-7} (\frac{300K}{T_\text{e}})^{0.6}$ \\
        \hline
        Na$^+$ & $\text{Na}^+ + \text{e}^{-} \rightarrow$ Na & $2.7 \times 10^{-12} (\frac{300K}{T_\text{e}})^{0.6}$ \\
        \hline
    \end{tabular}
    \label{tab:meteoric_ions}
\end{table}

\noindent \textbf{Equilibrium}.
Under the approximation of chemical equilibrium, which is valid in the lower ionospheres where chemical reaction times are smaller than transport times, we can assume an exact balance between production and recombination of each ion species.
To the reactions listed in Table \ref{tab:reactions}, we can add the reactions in Table \ref{tab:meteoric_ions}.
Now, if we consider all reactions, the electrical neutrality condition can be written as:
\begin{equation}
    [\text{e}^-] = [\text{H}_3^+] + [ \text{CH}_5^+] + [\text{Fe}^+] + [\text{Mg}^+] + [\text{Si}^+] + [\text{Na}^+]
\end{equation}

The resolution of the chemical network is given in the Supplementary Information.

\noindent \textbf{Conductances}.
The Pedersen and Hall conductances $\Sigma_\text{P}$, $\Sigma_\text{H}$ can be retrieved by height integration of  $\sigma_\text{P}$, $\sigma_\text{H}$: 
\begin{equation}
    \Sigma_\text{P, H} = \displaystyle \int_{0}^{+\infty} \sigma_\text{P, H}(z) dz
    \label{equ:Sigma}
\end{equation}

%%%%%%%%%%%%%%%%%%%%%
% 3 - conductivities
%%%%%%%%%%%%%%%%%%%%%

\section{Conductivity profiles and impact of meteoric ions: results}
\label{Conductivities}

In this section, we present the conductivity profiles obtained with our model. The model calculates the ion densities and the associated conductivities for a given flux of precipitating electrons of a given characteristic energy.
We start with a short overview of real estimated fluxes and their associated energy distribution at giant planets.
\begin{itemize}
    \item At Jupiter, \citet{Grodent2001} use a Maxwellian spectral distribution centered on characteristic energies of the order of a fraction of a MeV.
    \\
    \citet{Gustin2004} obtain a positive correlation between the mean precipitating electron energy (between 30 and 200 keV) and the total energy flux (between 2 and 30 mW/m$^2$).
    %(mW/m$^2$=erg/s/cm$^2$))
    \item At Saturn, \citet{Gustin2017} exhibit characteristic energies of a few keV and fluxes of a few mW/m$^2$.
    \item At Uranus, \citet{Waite1988} estimate a total precipitated energy flux of 0.9 mW/m$^2$ in the aurora with a characteristic energy of 10 keV (0.008 mW/m$^2$ averaged over the entire disk).
    \item At Neptune, only speculative fluxes and characteristic energies can be taken for now. Values similar to those of Uranus seem reasonable.
\end{itemize}

1 mW/m$^2$ seems to be a good value for a characteristic flux at Saturn, Uranus and Neptune, while 100 mW/m$^2$ is a good upper limit for the four planets.
In Section \ref{Conductances}, we will study the conductances induced by these two different fluxes, 1 mW/m$^2$ and 100 mW/m$^2$, over a spectrum of characteristic energy of precipitating electrons.
To study the conductivities profiles in this section, we test two different cases for each planet:
a 100 mW/m$^2$ flux with a characteristic energy of 100 keV, and a 1 mW/m$^2$ flux with a characteristic energy of 1 keV. This choice illustrates the two typical cases. The ``Jupiter" case with a high flux and a high characteristic energy and the ``other giants" case with a low flux and a low characteristic energy.
 
For each of the 8 configurations (4 planets and 2 cases for each planet), we have made two runs of the model:
\begin{itemize}
    \item A run with meteoric ions (w mi)
    \item A run without meteoric ions (w/o mi): meteoric ion production rates are zero
\end{itemize}

\begin{figure}[ht!]
    \centering
    \includegraphics[width=0.90\linewidth]{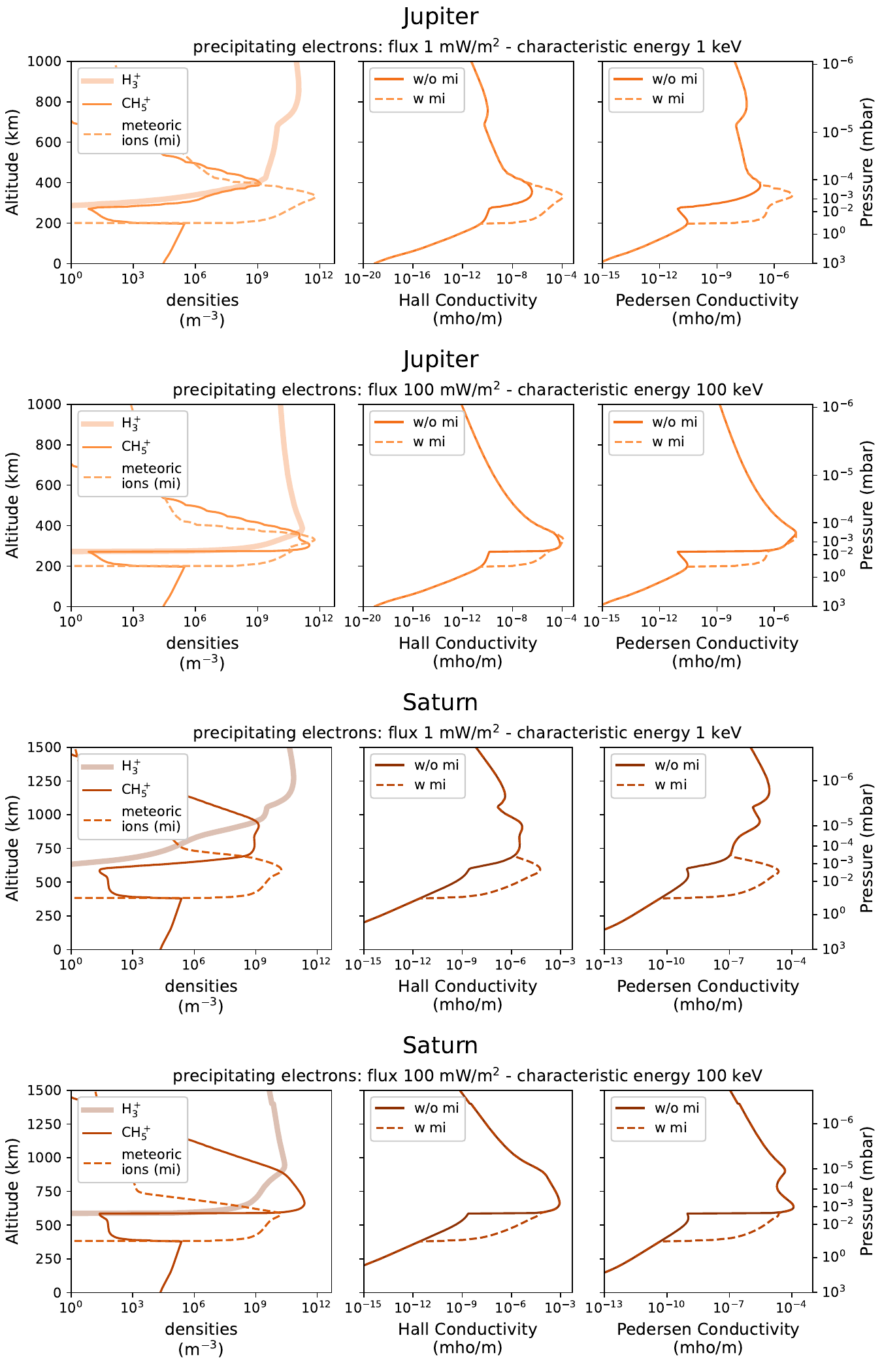}
\caption{
%%%\color{red}
Ion densities (first panel of each row) for a run with meteoric ions, and conductivities (second and third panels of each row) for runs with and without meteoric ions (w mi and w/o mi), in Jupiter and Saturn, for a 100mW/m$^2$ flux of precipitating electrons with two different characteristic energies: 1 keV (first and third rows) and 100 keV (second and fourth rows).
%%%\color{black}
}
\label{fig:J_S_conductivities}
\end{figure}

\begin{figure}[ht!]
    \centering
    \includegraphics[width=0.90\linewidth]{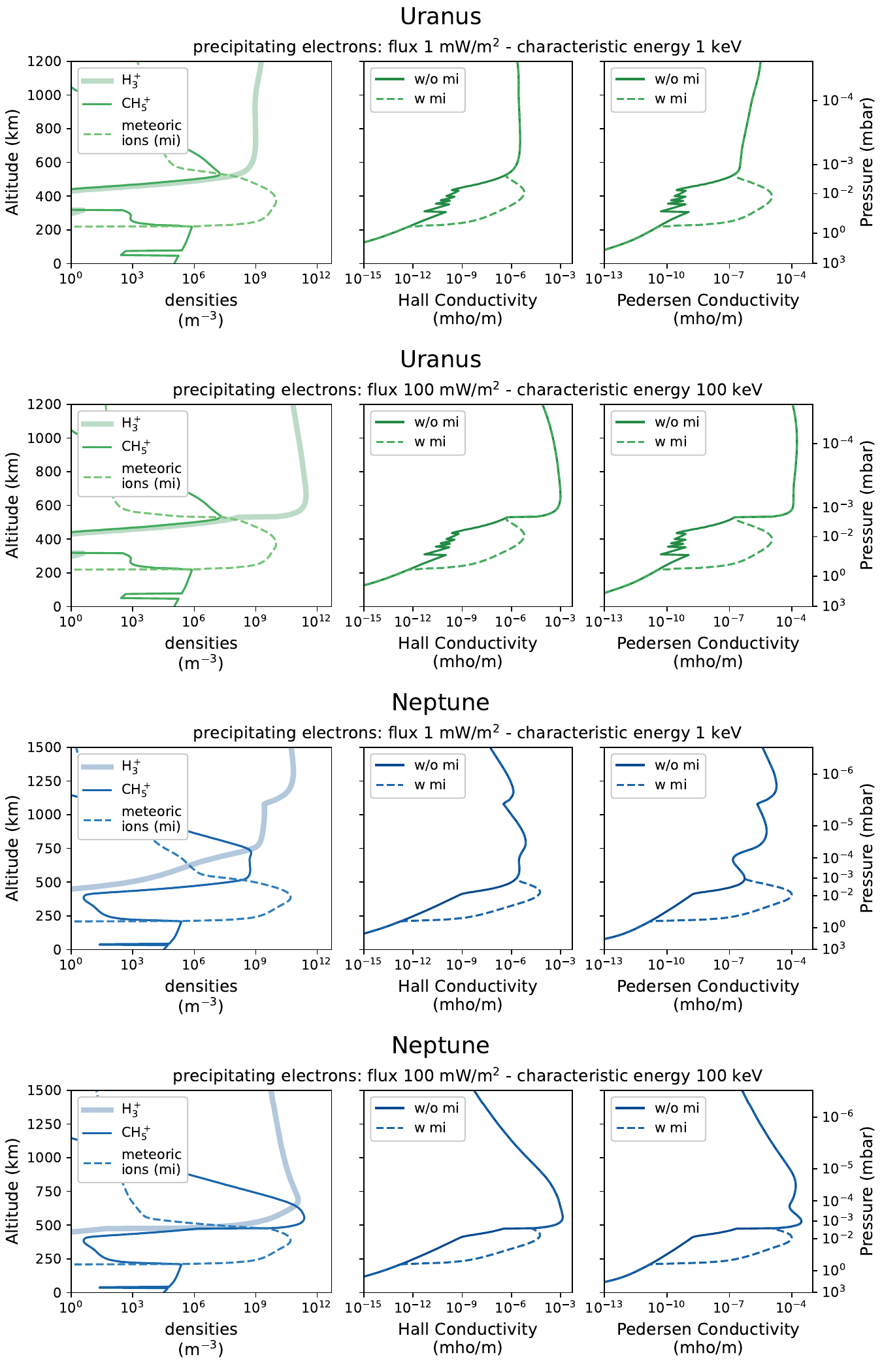}
\caption{Same as Figure \ref{fig:J_S_conductivities} for Uranus and Neptune}
\label{fig:U_N_conductivities}
\end{figure}

Figures \ref{fig:J_S_conductivities} and \ref{fig:U_N_conductivities} show the ion densities and associated conductivities calculated with these runs.
For the run with meteoric ions, we plot the ion densities in each first panel. The Hall and Pedersen conductivities associated with this case are plotted in dashed lines in the second and third panels of each Figure.
For the run without meteoric ions, we do not plot the ion densities, but we plot the Hall and Pedersen conductivities in solid lines in the second and third panels.

Conductivities are of the same order of magnitude at Saturn, Uranus and Neptune. At Jupiter, the conductivities are about 2 orders lower than in the three other planets for the same flux and characteristic energy, because of Jupiter's intense magnetic field.
We can see that the higher the characteristic energy of the precipitating electrons, the deeper (altitude-wise) the production peak of H$_3^+$.
At low altitudes, CH$_5^+$ ions are the most abundant ions in the four planets, in the lowest altitude ranges where conductivities are weak.
At higher altitudes, CH$_5^+$ ions have little effect in Jupiter and Uranus, since their density is always (or almost always) lower than the density of other species. At Saturn and Neptune, CH$_5^+$ ions do play a role and are the most abundant species at some altitude (about 800 km for Saturn and about 600 km for Neptune) for both cases (1 keV and 100 keV).
%%%\color{red}
There is a hole in CH$_5^+$ density profiles at roughly the same altitude as the meteoric ion peaks. Measurements would not be precise enough to check if such a hole is realistic.
It appears because of the structure of the chemical network: the production of meteoric ions increases a lot the production of electrons which leads to a higher rate of CH$_5^+$ recombination, diminishing their density.
Moreover, on each of the four giant planets, at the altitude range of 0-200 km (below this hole in CH$_5^+$ density), H$_3^+$ production is extremely low (almost zero), and consequently, CH$_5^+$ should also be very low. We could have chosen to set CH$_5^+$ density at zero whenever H$_3^+$ production is lower than an arbitrary threshold. But we preferred to keep it as calculated by the model, using it as an upper limit. Conductivities are very low in this altitude range of 0-200 km, which falls beyond the scope of this article.
%%%\color{black}

\begin{figure}[ht!]
    \centering
    \includegraphics[width=\textwidth]{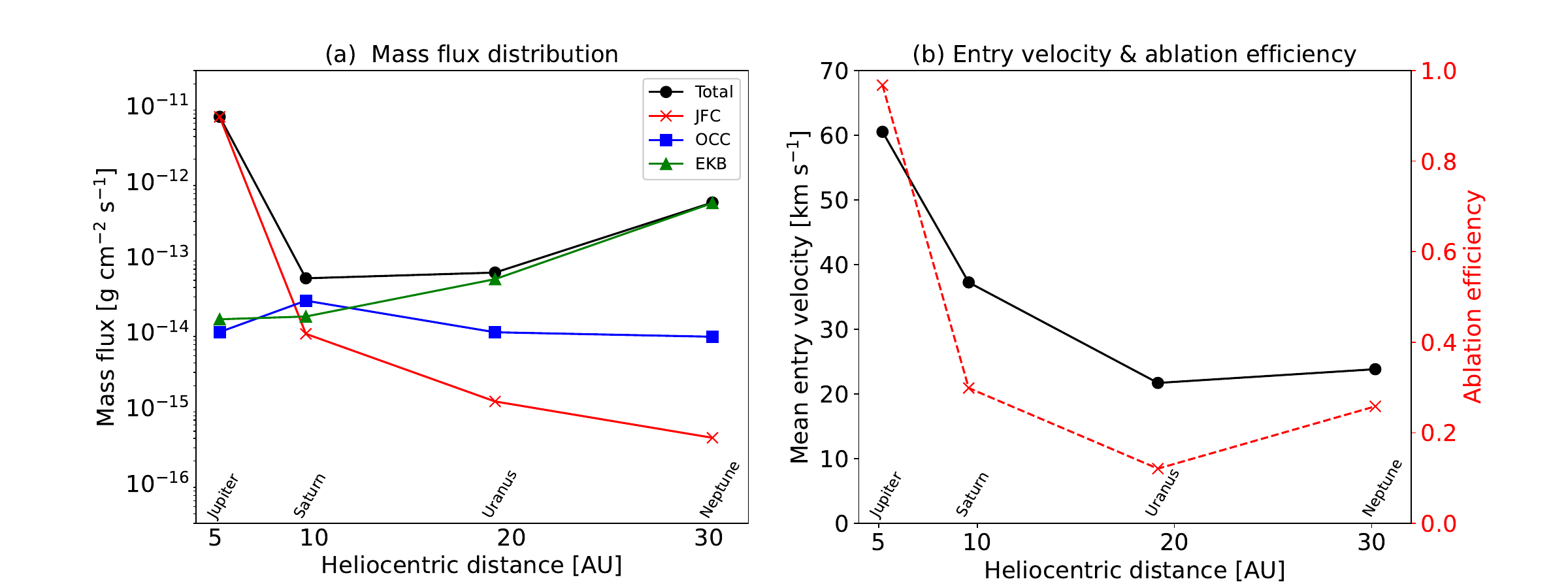}
    \caption{Mass flux distribution and mean entry velocity of meteoroids reaching each of the four giant planets.\\
    (a) Mass flux distribution of precipitating meteoroids on the four giant planets used in the simulation as a function of heliocentric distance [AU] for Jupiter-family comets (JFC, red), Oort-Cloud comets (OCC, blue), and Edgeworth–Kuiper belt (EKB, green). We used the mass and velocity distribution of meteoroids on each planet from \citet{Moses2017}.\\
    (b) Mean entry velocity of meteoroids from \citet{Moses2017}'s distribution and ablation efficiency of metallic species calculated by the meteoroid ablation model of \citet{Nakamura2022}.}
    \label{fig:mass_flux_entry_velocity}
\end{figure}

The production of ions from meteoric sources reaches its peak at a characteristic altitude for each planet.
If the energy of the precipitating electrons is high enough to penetrate and ionize the atmosphere at this characteristic altitude of meteoric ions production, then the contribution of H$_3^+$ ions can dilute the contribution of meteoric ions. 
However, precipitating electrons with lower energies do not penetrate as deep. In this case, meteoric ions become the most important ionization source at their characteristic altitude.
At Saturn, Uranus and Neptune, the peak of meteoric ion density is higher than 10$^{10}$ m$^{-3}$ in both cases. At Jupiter, this peak reaches 10$^{12}$ m$^{-3}$. This is because the mass flux and mean velocity of meteoroids precipitating into Jupiter's atmosphere are higher (because of its mass) than at the three other planets (see Figure \ref{fig:mass_flux_entry_velocity}). The total mass flux at Jupiter is about two orders of magnitude higher than at the three other planets. The ablation efficiency is proportional to the speed to the power of three. Mean entry velocities are higher at Jupiter (see Figure \ref{fig:mass_flux_entry_velocity}) than at the three other giant planets.
This induces a high ablation efficiency at Jupiter (almost 100\%, against $\sim$20\% for the three others).

Let us now focus on realistic cases for each planet, i.e [100 mW/m$^2$, 100 keV] for Jupiter and [1 mW/m$^2$, 1 keV] for the three other giant planets.
At Jupiter, meteoric ions add a significant contribution to Hall and Pedersen's conductivities. This contribution is of the same order as the contribution of H$_3^+$.
At Saturn, Uranus, and Neptune, the maximum density of meteoric ions is higher than the maximum densities of all other species, and meteoric ions thus contribute to Hall and Pedersen's conductivities.
At Jupiter, the production of meteoric ions occurs at similar altitudes as the production of H$_3^+$ and CH$_5^+$.
At the three other planets, the production of H$_3^+$ and CH$_5^+$ occurs higher than the production of meteoric ions. Conductivities at these planets are supported by different sources depending on the altitude.
Indeed, characteristic altitudes seem to play an important role and are investigated in more detail in Section \ref{Altitudes}.
Before that, we study the impact of meteoric ions on height-integrated conductances.

\FloatBarrier

\newpage

%%%%%%%%%%%%%%%%%%%%%
% 4 - conductances
%%%%%%%%%%%%%%%%%%%%%

\section{Impact of meteoric ions on conductances: results}
\label{Conductances}

In this section, we study the conductances calculated for 2 different fluxes of precipitating electrons (1 mW/m$^2$ and 100 mW/m$^2$) over a spectrum of characteristic energies extending from 0.1 to 10$^4$ keV.

\begin{figure}[ht!]
    \centering
    \includegraphics[width=\linewidth]{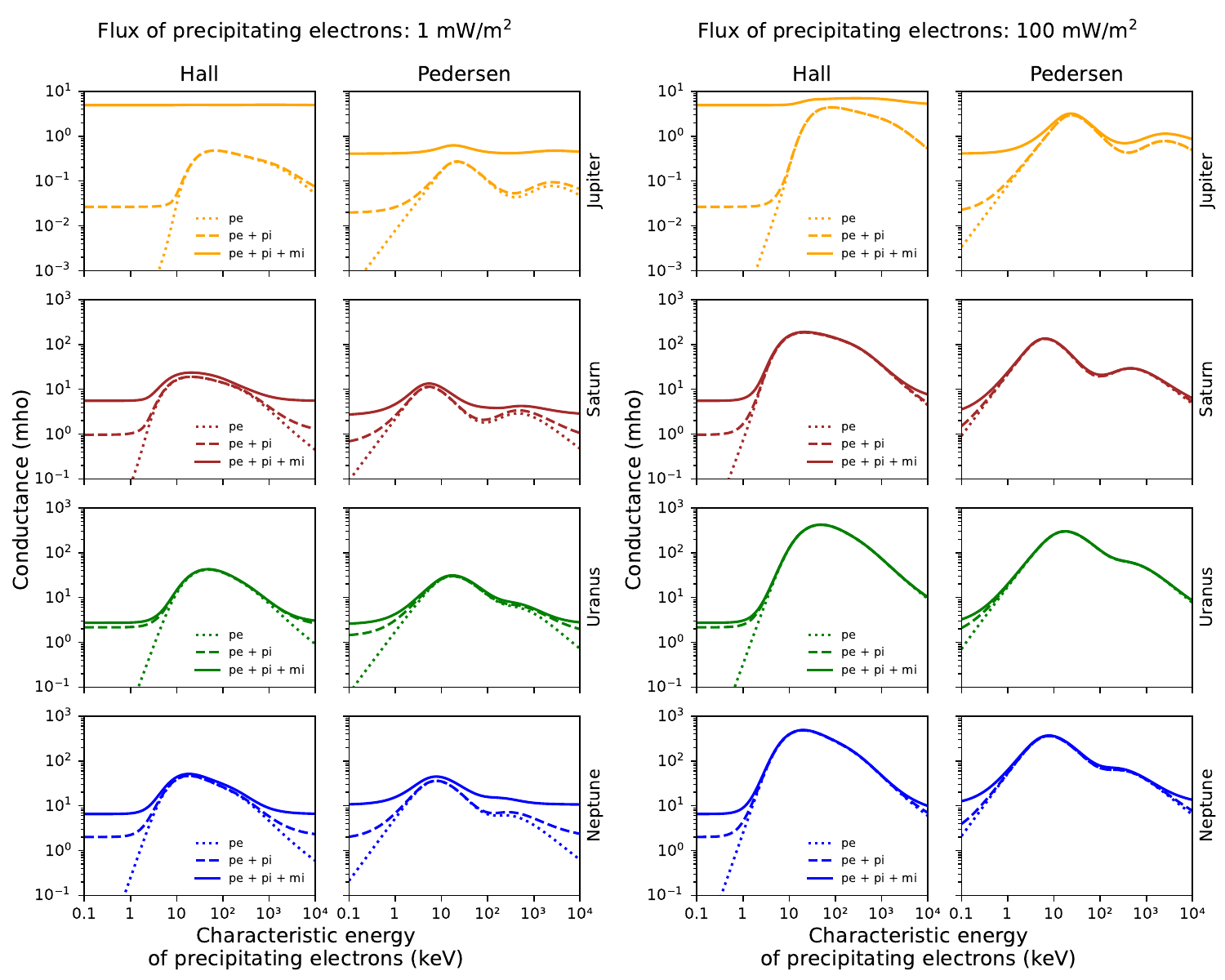}
    \caption{Hall and Pedersen conductances calculated for the four giant planets with the 3 main ionization sources added one by one: precipitating electrons (pe), photoionization (pi), meteoric ions (mi).\\
    Two different precipitating electron fluxes, 1 mW/m$^2$ in the first and second columns and 100 mW/m$^2$ in the third and fourth columns, and a range of characteristic energy (x-axis) from 0.1 keV to 10$^4$ keV are tested.}
    \label{fig:8_plots}
\end{figure}

\begin{figure}[ht!]
    \centering
    \includegraphics[width=\linewidth]{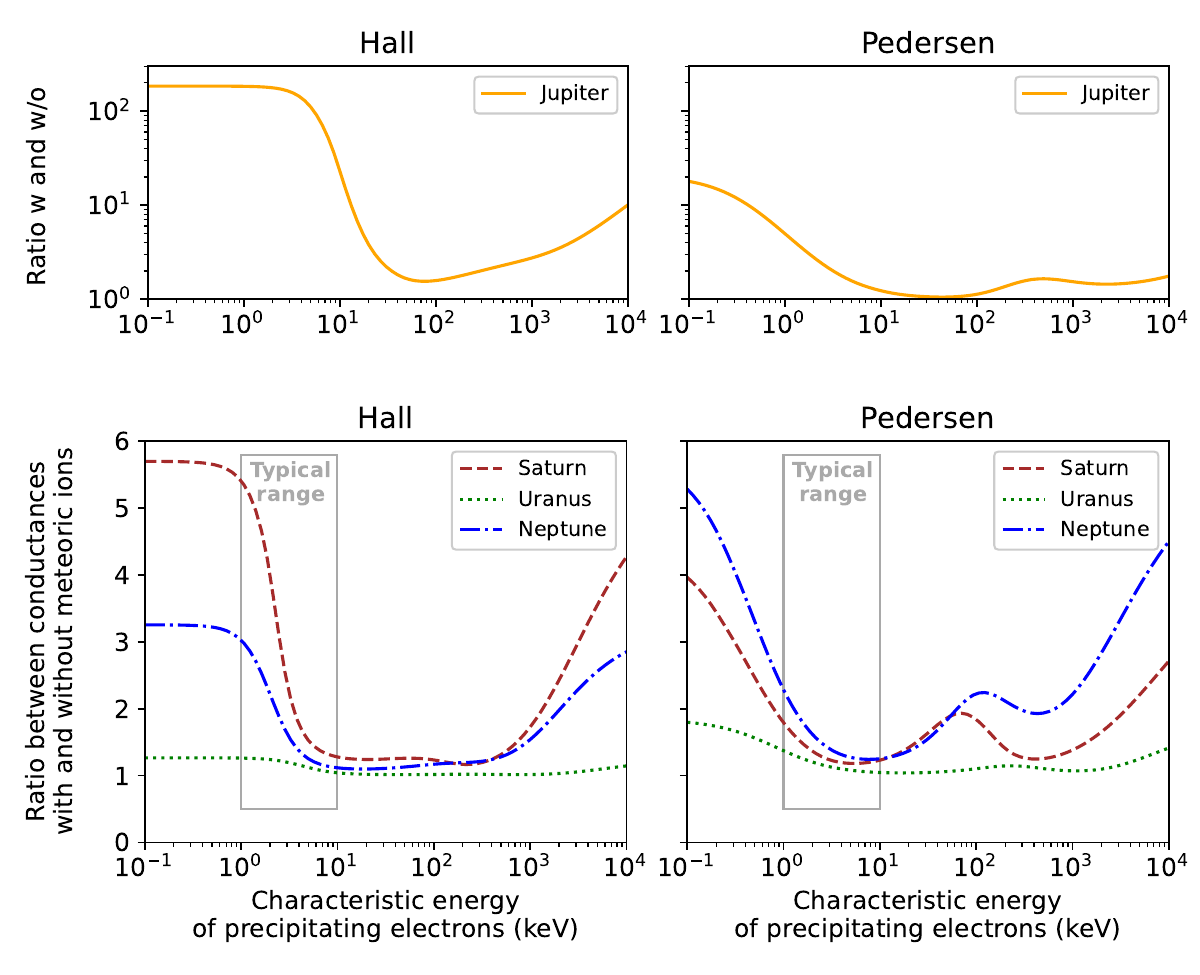}
    \caption{Ratio between conductances with contribution (w) of meteoric ions and conductances without contribution (w/o) of meteoric ions.
    The flux of precipitating electrons is 100 mW/m$^2$ for Jupiter (first row) and 1 mW/m$^2$ for Saturn, Uranus and Neptune (second row).}
    \label{fig:ratios}
\end{figure}

To calculate the conductances, we integrate the conductivities along the altitude (see equation \ref{equ:Sigma}).
We plot the conductances calculated over the energy spectrum at the four giant planets in Figure \ref{fig:8_plots}, for 3 cases. In the first case, we consider only ionization by precipitating electrons (pe). In the second case, we add photoionization (pi). We denote this case by ``pe+pi". It corresponds to the case without meteoric ions (w/o mi) in Figures \ref{fig:J_S_conductivities} and \ref{fig:U_N_conductivities}. In the third case, we finally add meteoric ions. We denote this case by ``pe+pi+mi". It corresponds to the case with meteoric ions (w mi) in Figures \ref{fig:J_S_conductivities} and \ref{fig:U_N_conductivities}.
In addition to Figure \ref{fig:8_plots}, the ratios between the cases with (pe+pi+mi) or without (pe+pi) the contribution of meteoric ions are plotted in Figure \ref{fig:ratios}.

\par
The contribution of meteoric ions to both Hall and Pedersen conductances is found to be significant at Jupiter, whatever the characteristic energy of precipitating electrons. Results from \citet{Nakamura2022} are recovered.
Looking at the results for a 100 mW/m$^2$ flux at Jupiter (Figure \ref{fig:8_plots}b), this contribution is significant for the Hall conductance at all characteristic energies.
For the Pedersen conductance, the contribution of meteoric ions is reduced when characteristic energies of precipitating electrons are around 50 keV (see Figure \ref{fig:ratios}).

\par
At Saturn, Uranus and Neptune, this contribution becomes significant when the characteristic energy of precipitating electrons is lower than a few keV.
For precipitating electrons of 1 keV, the contribution of meteoric ions multiplies the Hall conductance by 6 at Saturn, by 1.5 at Uranus and by 3 at Neptune, and the Pedersen conductance by 6 at Saturn, by 1.5 at Uranus and by 3 at Neptune.
The range of 1-10 keV is the typical energy range of precipitating electrons at these three planets (at least at Saturn and Uranus, and probably also at Neptune).
We can see in Figure \ref{fig:8_plots} that this range is critical.
For a characteristic energy of 10 keV, precipitating electrons are energetic enough to be the main contribution to both Hall and Pedersen conductances and to drastically reduce the contribution of meteoric ions (see in particular Figure \ref{fig:ratios}).
For a characteristic energy of 1 keV, precipitating electrons are less energetic. Meteoric ions become the main contribution to both Hall and Pedersen conductances.
If the energy of precipitating electrons does not exceed a few keVs, the contribution of meteoric ions to conductances is not negligible.

\par
The plots in the previous section did not highlight the contribution of photoionization.
Its impact is now displayed.
By looking at ``pe+pi" cases when energy is 0.1 keV (extremely low energy, no impact of precipitating electrons), we can see photoionization's contribution to conductance.
It is about 10$^{-2}$ mho at Jupiter and a few mhos at the three other giant planets.
When adding meteoric sources (case pe+pi+mi) the contribution of photoionization becomes systematically diluted at Jupiter, Saturn, and Neptune.
At Uranus, the ablation efficiency is lower, meteoric sources are reduced, and compete with the contribution of photoionization to conductances, which are of the same order of magnitude (about 1 mho each).
Unlike at Earth, where photoionization is dominant, the contribution of photoionization at giant planets is smaller and decreases sharply with heliocentric distance.

\par
Additional remarks can be made.
Although no precipitating electrons exist at such high energies, we plot the range [10$^3$-10$^4$] keV to better capture the dependence of conductances on the energy spectrum. In this range, the contribution of meteoric ions would also theoretically be significant.
The intensity of the fluxes (1 mW/m$^2$ or 100 mW/m$^2$) of precipitating electrons regulates the intensity of conductances, but only in the energy range within which this ionization process is the most important: the larger the flux, the larger the conductance.
We can also see that the energy at which conductances reach their maximum is larger at Jupiter than at the three other planets, independently of the absolute value of precipitation energy fluxes.

With our simplified model, we show that the contribution of meteoric ions to height-integrated conductances at the four giant planets (and not only Jupiter) could be non-negligible when the energy of precipitating electrons is lower than a few keVs and should be carefully taken into account when using ionospheric conductance values for these planets.

%%%%%%%%%%%%%%%%%%%%%
% 5 - conductive layer
%%%%%%%%%%%%%%%%%%%%%

\section{Altitudes of the conductive layer and of the meteoric ion layer, and implications for magnetosphere-ionosphere coupling: results}
\label{Altitudes}

In this section, we discuss the main differences in altitude-dependent variations in conductivities between the giant planets and their implications for magnetosphere-ionosphere coupling.

Looking back at equations \ref{sigma_P} and \ref{sigma_H}, we can see that the intensity of conductivities (and conductances) is governed by both:
\begin{itemize}
    \item the $\dfrac{n_\text{e} e}{B}$ and $\dfrac{n_\text{i} e}{B}$ terms, that are inversely proportional to the magnetic field
    \item the $\dfrac{\left|\omega_\text{ce}\right|\nu_\text{en}}{\nu_\text{en}^2 + \omega_\text{ce}^2}$, $\dfrac{\left|\omega_\text{ci}\right|\nu_\text{in}}{\nu_\text{in}^2 + \omega_\text{ci}^2}$, $\dfrac{\omega_\text{ce}^2}{\nu_\text{en}^2 + \omega_\text{ce}^2}$, $\dfrac{\omega_\text{ci}^2}{\nu_\text{in}^2 + \omega_\text{ci}^2}$ terms, that need to be close to 1 for the conductivity to be significant
\end{itemize}

\begin{figure}[ht!]
    \centering
    \includegraphics[width=\linewidth]{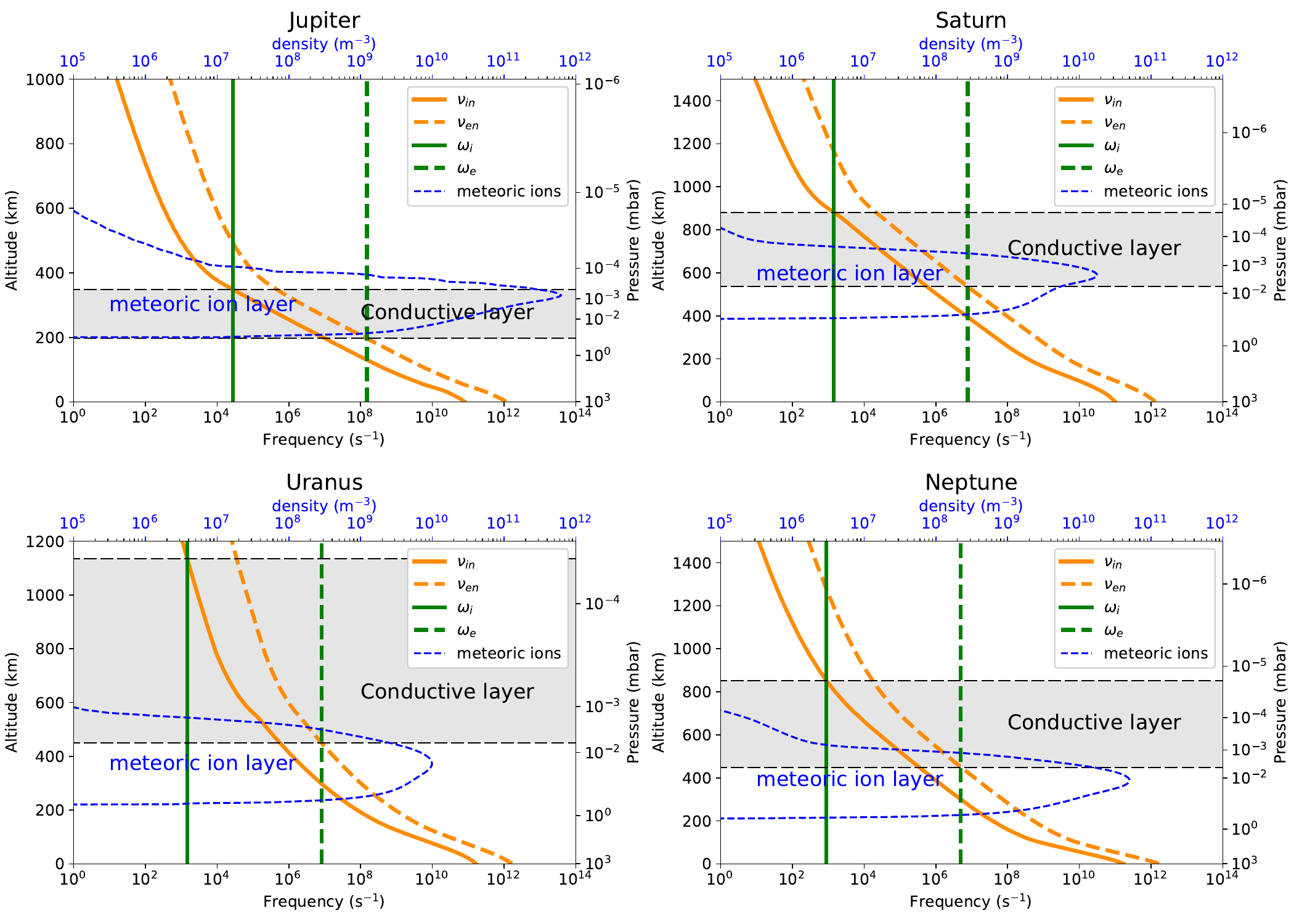}
    \caption{Comparison of the altitudes of the conductive layer and of the meteoric ion layer.\\
    The conductive layer is defined by the ratios between the cyclotron frequencies and the collision frequencies with neutrals: $\omega_\text{ce}$/$\nu_\text{en}$ and $\omega_\text{ci}$/$\nu_\text{in}$ ($\nu_\text{in}$ and $\omega_\text{i}$ are given for H$_3^+$ ions). The horizontal dashed black lines show where these ratios are equal to 1. The conductive layer is then mainly located between these two lines.\\
    The meteoric ion layer is defined as the altitude range where most of the meteoric ions are produced.}
\label{fig:layers}
\end{figure}

The latter are close to 1 when $\omega_\text{ce}$ and $\nu_\text{en}$, or $\omega_\text{ci}$ and $\nu_\text{in}$, are of the same order of magnitude. The collision frequencies $\nu_\text{en}$ and $\nu_\text{in}$ are proportional to the neutral density and exponentially increase with depth. The cyclotron frequencies $\omega_\text{ce}$ and $\omega_\text{ci}$ of electrons and ions are inversely proportional to the magnetic field intensity, which varies very little across the ionospheric layer.
In Figure \ref{fig:layers}, we plot, as functions of altitude, $\nu_\text{en}$ and $\omega_\text{e}$, $\nu_\text{in}$ and $\omega_\text{i}$ for H$_3^+$ ions. $\nu_\text{in}$ of other ions is comprised between $0.8\times \nu_\text{in}$(H$_3^+$) and $\nu_\text{in}$(H$_3^+$), because $\nu_\text{in}$ is proportional to $1/\sqrt{\mu_\text{a}}$ (see equation \ref{i_n_collision_frequency}) - $\mu_\text{a}$ being the reduced mass comprised between 1.2 and 1.93 AMU.
$\omega_\text{i}$(H$_3^+$) is an upper limit for the cyclotron frequency of other ions. The upper black horizontal line, plotted at the altitude where $\nu_\text{in}$ and $\omega_\text{i}$ are equal, can be then considered as a lower limit to define the upper part of the conductive layer, which only slightly extends above it.

\begin{figure}[ht!]
    \centering
    \includegraphics[width=0.7\linewidth]{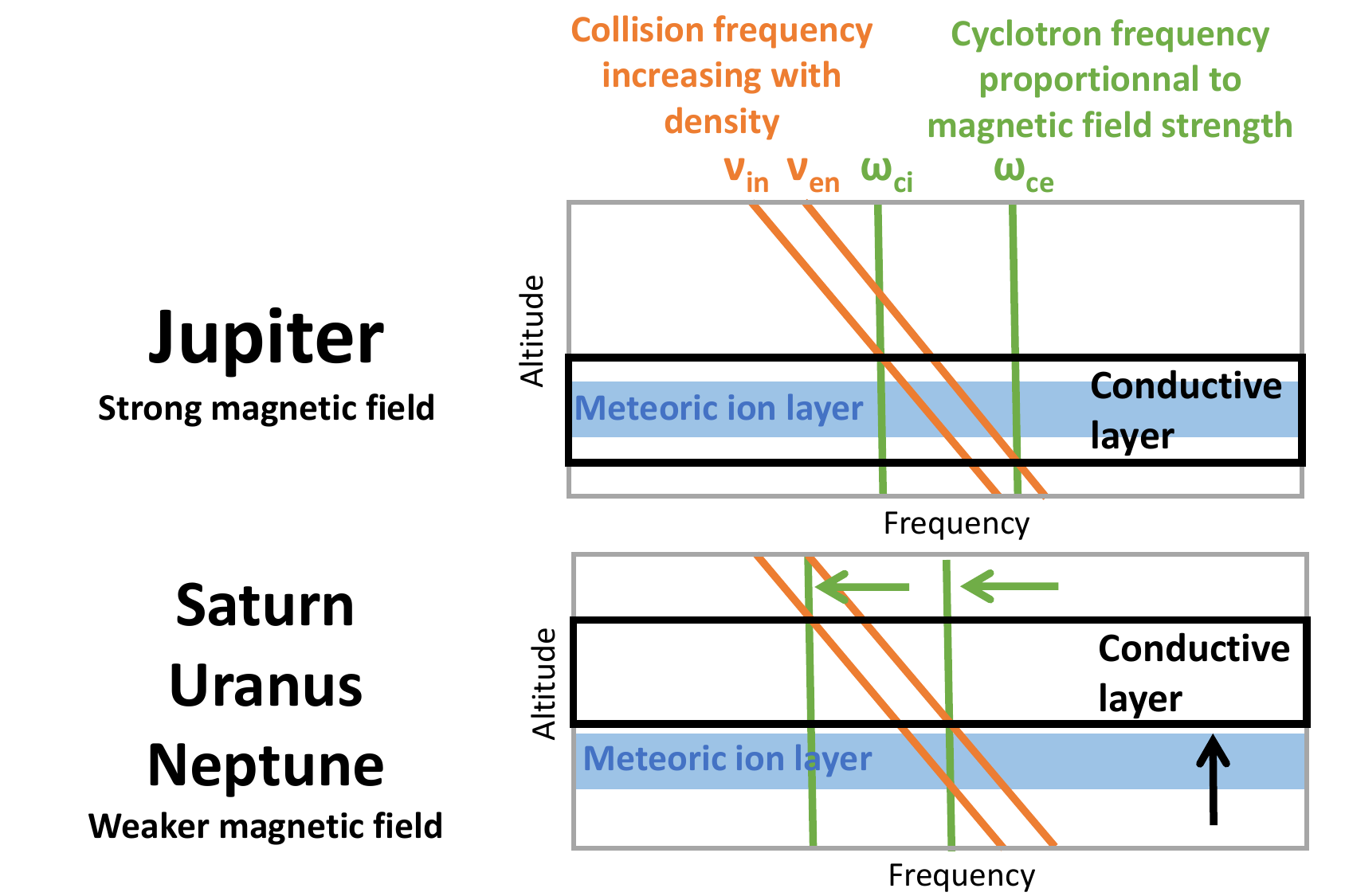}
    \caption{
    Schematic illustration of the conductive layers (black shadings) and meteoric ion layers (blue rectangles) at the four giant planets. The conductive layer is the region where the conditions  $\omega_\text{ci}<\nu_\text{in}$ and  $\omega_\text{ce}>\nu_\text{en}$  are satisfied. \\
    $\omega_\text{ci}$ and $\omega_\text{ce}$ are the cyclotron frequencies of ions and electrons, $\nu_\text{in}$ and $\nu_\text{en}$ are collision frequencies of ions and electrons with neutrals. \\
    This Figure is inspired from Figure 8 in \citet{Nakamura2022}}
    \label{fig:layers_frequency_altitude}
\end{figure}

It is in this upper part of the conductive layer that the MI coupling is strongest. Moving down from the top to the bottom of the atmosphere, this upper part corresponds to the first peak of the Pedersen conductivity and, therefore, to the altitudes where most of the Pedersen currents connecting upward and downward field-aligned currents flow. 
In Figure \ref{fig:layers}, we can see that, at Jupiter, this upper black horizontal line coincides with the altitude of the maximum meteoric ion density. On the contrary, at Saturn, Uranus and Neptune, the upper black horizontal line is clearly higher by a few hundred km (300 km for Saturn, 700 km for Uranus, 500 km for Neptune) than the altitude of the maximum meteoric ion density: the weaker magnetic field of these three planets reduces cyclotron frequencies and shifts the conductive layer to higher altitudes. This effect is summarized in Figure \ref{fig:layers_frequency_altitude}, which shows that the stronger the magnetic field, the deeper (altitude-wise) the conductive layer.

\FloatBarrier

%%%%%%%%%%%%%%%%%%%%%
% 6 - Discussion
%%%%%%%%%%%%%%%%%%%%%

\section{Discussion: limitations and open questions}
\label{Discussion}

Our model relies on the superposition of 3 different models that calculate ionization rates from 3 different sources:
\begin{itemize}
    \item photons \citep{Richards1994}
    \item electrons \citep{Hiraki2008}
    \item meteoroids \citep{Nakamura2022}
\end{itemize}

The three models are applied assuming that atmospheres of giant planets are H$_2$-dominated, thus allowing us to consider them as huge reservoirs of H$_2$.
Photons and electrons ionize H$_2$. Meteoroids are ablated through friction with the atmosphere (so mainly H$_2$).
Our model assumes photo-chemical equilibrium, which is a strong approximation but gives an overview of steady-state.
In addition to H$_2$-associated ions and meteoric ions, only CH$_5^+$ is considered because it is the main product of H$_3^+$ with other molecules. A more complex chemical scheme implying a broader list of reactions could be implemented.

\noindent \textbf{Precipitating electrons.} We could run our model using more realistic fluxes of precipitating electrons, i.e. with energy spectrum dependence.
Because this can only be done at Jupiter and Saturn (there is a lack of information on such fluxes at ice giants), we chose to limit ourselves to a more simple and comparative approach.
The dependence of conductances on the energy of precipitating electrons has been shown in the section \ref{Conductances}. 

\noindent \textbf{Meteoric ions.}
We assumed that the ionization probability equals 0.4 at the four planets. This value has been estimated at Jupiter \citep{Kim2001}. This value depends on the entry velocity of meteoroids, could be smaller at Saturn, Uranus, and Neptune than at Jupiter, and may also differ for each element \citep{Vondrak2008}.
Meteoric ions have long lifetimes (more than 100 Jovian days, \citet{Nakamura2022}).
These are long enough lifetimes to maintain their density even if the meteoroid influx is set to be asymmetric in local time.

\noindent \textbf{Longitude and latitude.} Our model is a 1D model and focuses on auroral regions by taking electron fluxes from these regions. Variations with latitude and longitude should be investigated further in detail.

%%%%%%%%%%%%%%%%%%%%%
% 7 - conclusions
%%%%%%%%%%%%%%%%%%%%%

\section{Conclusions}

Using a generic ionospheric model applied to the four giant planets of our Solar System - Jupiter, Saturn, Uranus, and Neptune - we have estimated and compared the ionospheric conductivities and conductances at these planets produced by photoionization, electron precipitation and meteoric sources. In addition to the main ions H$_3^+$ and CH$_5^+$, this model includes ions produced by the precipitation and ablation of meteoroids in the atmosphere.
Photons, electrons, meteoroids: 3 very different types of particles that interact with an H$_2$-atmosphere.
Our main conclusions are: 
\begin{itemize}
    \item Photoionization plays little role in the auroral regions, compared to the two other ionization processes, at the four planets.
    \item With our simplified approach, the conclusions of \citet{Nakamura2022} are well reproduced at Jupiter: the contribution of meteoric ions to the conductances is non-negligible.
    \item At Saturn, Uranus and Neptune, the contribution of meteoric ions to conductances could also be non-negligible with respect to the ionization by precipitating electrons of characteristic energy lower than a few keV.
    \item The 1-10 keV range for characteristic energy of precipitating electrons is the typical range at these three planets. In this range, the two ionization processes compete. At 10 keV (and above), precipitating electrons are energetic enough to be the main ionization source. However, at 1 keV, the contribution of meteoric ions is no longer diluted and should be taken into account.
    \item The top of the ionospheric conductive layer - and especially its upper part, which couples magnetosphere and ionosphere - is deeper at Jupiter than at Saturn, Uranus and Neptune because of the stronger magnetic field of Jupiter.
    \item The meteoric ion layer, where meteoric ions are mostly produced and have their highest density, coincides with the top of this conductive layer in Jupiter, contributing significantly to Hall and Pedersen conductivities, as already demonstrated in \citet{Nakamura2022}.
    \item Conversely, at Saturn, Uranus and Neptune, because of their weaker magnetic field, the conductive layer is higher than the layer where meteoric ions are mainly produced. The contribution of these ions to magnetosphere-ionosphere coupling is thus limited.
    \item The parameter space of our ionospheric conductivity model, which includes both magnetic field intensity (directly), planetary gravity and atmosphere composition (via the atmospheric profiles) can easily cover or be extended to the atmospheres of Jupiter-like and Neptune-like exoplanets, thus allowing for predictions of the role and importance of magnetosphere-ionosphere coupling at these planets. 
\end{itemize}

\section*{Data Availability Statement}

The modeling data supporting the figures presented in this paper are available at \url{https://doi.org/10.5281/zenodo.13992317}.

\section*{Acknowledgement}

The authors wish to thank Prof. Jiuhou Lei of the University of Science and technology of China for his kind support in the evaluation of the relative importances of solar and stellar ionisation rates in our model. 

N. Clément acknowledges the support of the French Agence Nationale de la Recherche (ANR), under grant ANR-20-CE49-0009 (project SOUND).

M. BLanc wishes to express his gratitude to CNES for its support to his participation in the Juno and JUICE missions, which has inspired and motivated this study.

Y. X. Wang was supported by NNSFC Grants 42304189, Pandeng Program of National Space Science Center, Chinese Academy of Sciences.

\bibliography{reference}

\end{document}